\documentclass[draft,jgrga]{agu2001}
\usepackage{graphicx}
\authorrunninghead{V\"{O}R\"{O}S ET AL.}
\titlerunninghead{Reconnection associated multi-scale fluctuations}
\authoraddr{Z. V\"{o}r\"{o}s,
Space Research Institute, Austrian Academy of Sciences, Schmiedlstrasse 6, 8042 Graz, Austria.
(zoltan.voeroes@oeaw.ac.at)}

\begin{document}

%% ------------------------------------------------------------------------ %%
%
%  ENABLE IMAGE DISPLAY WHILE USING DRAFT MODE
%
%% ------------------------------------------------------------------------ %%
%
% Uncomment the following code (as well as \usepackage{graphicx} above)
% if you need to include images in draft mode
%\setkeys{Gin}{draft=false}
%
% PLEASE NOTE: WHEN YOU SUBMIT YOUR LATEX FILE TO GEMS, COMMENT OUT ANY COMMANDS
% THAT INCLUDE GRAPHICS.
% (See FIGURES section near the end of the file)
%

%% ------------------------------------------------------------------------ %%
%
%  TITLE
%
%% ------------------------------------------------------------------------ %%

\title{Study of reconnection-associated multi-scale fluctuations with
Cluster and Double Star}
%

%% ------------------------------------------------------------------------ %%
%
%  AUTHORS AND AFFILIATIONS - 3 methods
%
%% ------------------------------------------------------------------------ %%

% Method 1 (for all journals, except Reviews of Geophysics, which
% should use method 3):
% For three or fewer author/affiliation blocks, use \author{} and \affil{}

%\author{Z. V\"{o}r\"{o}s}
%\affil{Space Research Institute, Austrian Academy of Sciences, Graz, Austria}
% ---------------
% Method 2 (for all journals, except Reviews of Geophysics, which
% should use method 3):
% For more than three author/affiliation blocks,
% use \author{\altaffilmark{}} and \altaffiltext{}
% \altaffilmark will produce footnote;
% matching altaffiltext will appear at bottom of page.
% May use \\ to start a new line.
\smallskip
\smallskip
\smallskip
  \authors{Z. V\"{o}r\"{o}s, \altaffilmark{1,2}
R. Nakamura, \altaffilmark{1} V. Sergeev, \altaffilmark{3}
W.Baumjohann, \altaffilmark{1} A. Runov, \altaffilmark{1} T.L.
Zhang, \altaffilmark{1} M. Volwerk, \altaffilmark{1,4} T. Takada,
\altaffilmark{5} D. Jankovi\v{c}ov\'{a}, \altaffilmark{2} E. Lucek,
\altaffilmark{6} and H. R\`eme \altaffilmark{7} }

\altaffiltext{1} {Space Research Institute, Austrian Academy of
Sciences, Graz, Austria.} \altaffiltext{2}{Institute of Atmospheric
Physics, Prague, Czech Republic.} \altaffiltext{3}{St. Petersburg
University, Russia.} \altaffiltext{4}{Max-Planck-Institut f\"ur
extraterrestrische Physik, Garching, Germany.}
\altaffiltext{5}{Japan Aerospace Exploration Agency, Sagamihara,
Japan.} \altaffiltext{6}{Imperial College, London, UK.}
\altaffiltext{7}{CESR, Toulouse, France.}
%% ------------------------------------------------------------------------ %%
%
%  ABSTRACT
%
%% ------------------------------------------------------------------------ %%

% Do NOT include any \begin...\end commands within
% the body of the abstract.

\begin{abstract}
The objective of the paper is to asses the specific spectral scaling
properties of magnetic reconnection associated
fluctuations/turbulence at the Earthward and tailward outflow
regions observed simultaneously by the Cluster and Double Star
(TC-2) spacecraft on September 26, 2005. Systematic comparisons of
spectral characteristics, including variance anisotropy and
scale-dependent spectral anisotropy features in wave vector space
were possible due to the well-documented reconnection events,
occurring between the positions of Cluster (X = -14--16 $R_e$) and
TC-2 (X = -6.6 $R_e$). Another factor of key importance is that the
magnetometers on the spacecraft are similar. The comparisons provide
further evidence for asymmetry of physical processes in
Earthward/tailward reconnection outflow regions. Variance anisotropy
and spectral anisotropy angles estimated from the multi-scale
magnetic fluctuations in the tailward outflow region show features
which are characteristic for magnetohydrodynamic cascading
turbulence in the presence of a local mean magnetic field. The
multi-scale magnetic fluctuations in the Earthward outflow region
are exhibiting more power, lack of variance and scale dependent
anisotropies, but also having larger anisotropy angles. In this
region the magnetic field is more dipolar, the main processes
driving turbulence are flow breaking/mixing, perhaps combined with
turbulence ageing and non-cascade related multi-scale energy
sources.

\end{abstract}

%% ------------------------------------------------------------------------ %%
%
%  BEGIN ARTICLE
%
%% ------------------------------------------------------------------------ %%

% The body of the article must start with a \begin{article} command,
% and an \end{article} command must be placed at the end of the file,
% before \end{document}.
%
% If using draft mode \end{article} must follow the references section.

\begin{article}

%% ------------------------------------------------------------------------ %%
%
%  TEXT
%
%% ------------------------------------------------------------------------ %%
 \begin{figure*}
\noindent\includegraphics[width=30pc]{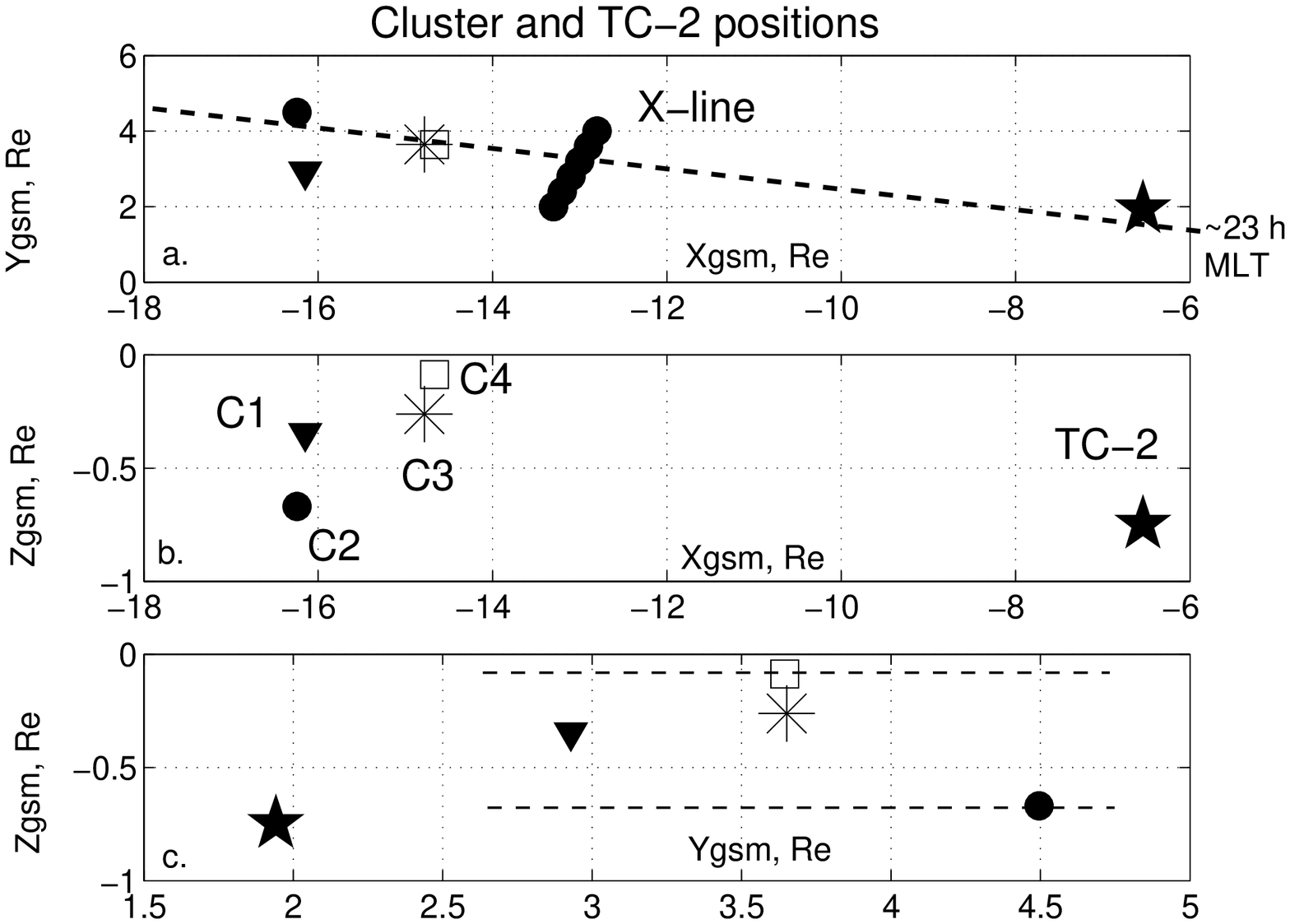}
 \caption{GSM positions of Cluster and TC-2 spacecraft; (a) X-Y plane; the
 approximate position of the magnetic X line is depicted; the Cluster - X-line - TC-2
 constellation is aligned along the same meridian (~23 MLT); (b) X-Z plane;
 (c) Y-Z plane; the horizontal dashed lines correspond to the GSM Z extent
 of reconnection outflow at Cluster positions around 0942 UT.}
 \end{figure*}

 \begin{figure*}
\noindent\includegraphics[width=30pc]{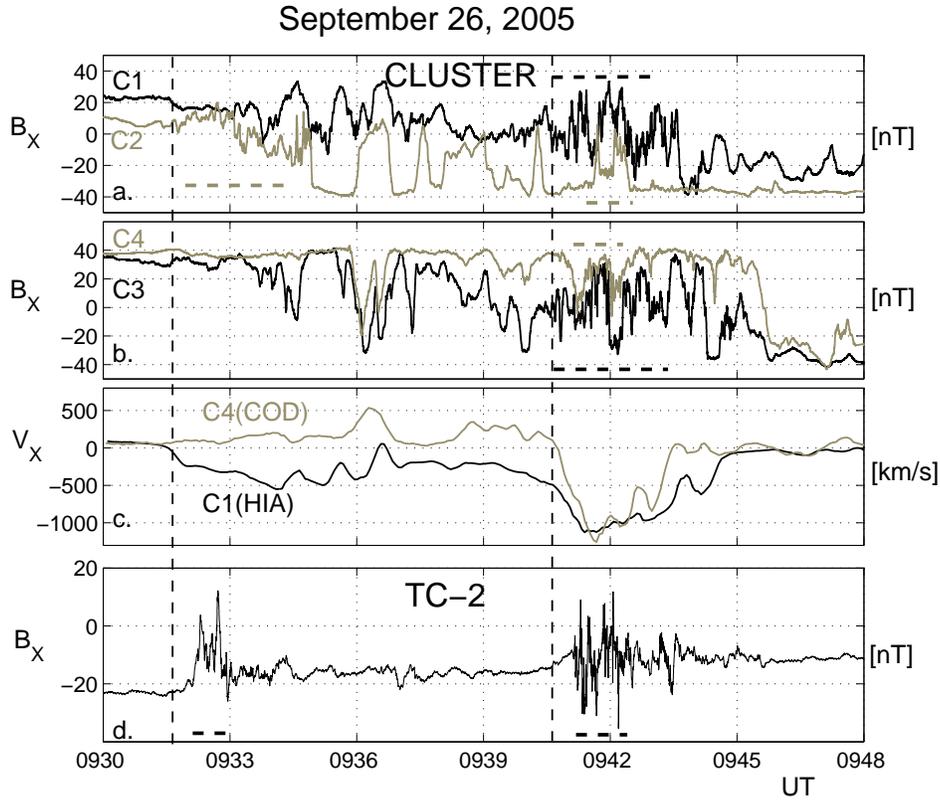}
 \caption{Cluster and TC-2 measurements. (a-b) $B_X$ magnetic components
 from Cluster; (c) $V_X$ components of the bulk velocity
 from C1 and C4; (d)$B_X$ magnetic component from TC-2. The vertical dashed
 lines show the beginning of multi-scale fluctuations in Cluster;
 the horizontal dashed lines show the length of particular multi-scale
 fluctuation intervals in Cluster and TC-2.}
 \end{figure*}

 \begin{figure*}
\noindent\includegraphics[width=30pc]{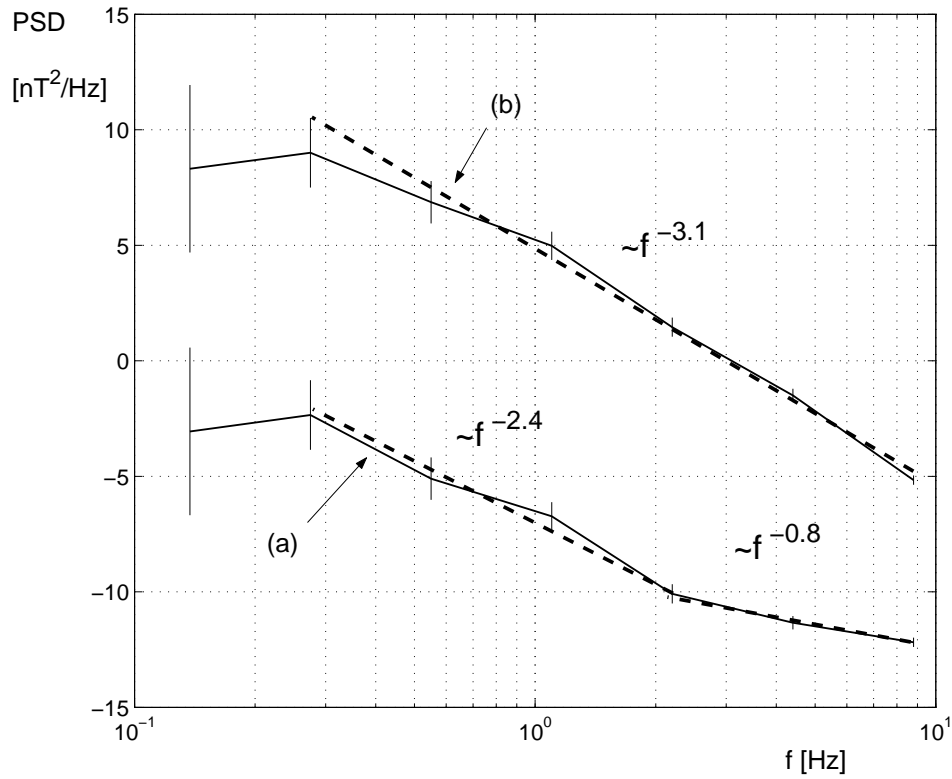}
 \caption{Comparison of power spectra estimated from TC-2 data;
(a) during a quiet interval; (b) during reconnection associated
fluctuations.}
 \end{figure*}

 \begin{figure}
\noindent\includegraphics[width=30pc]{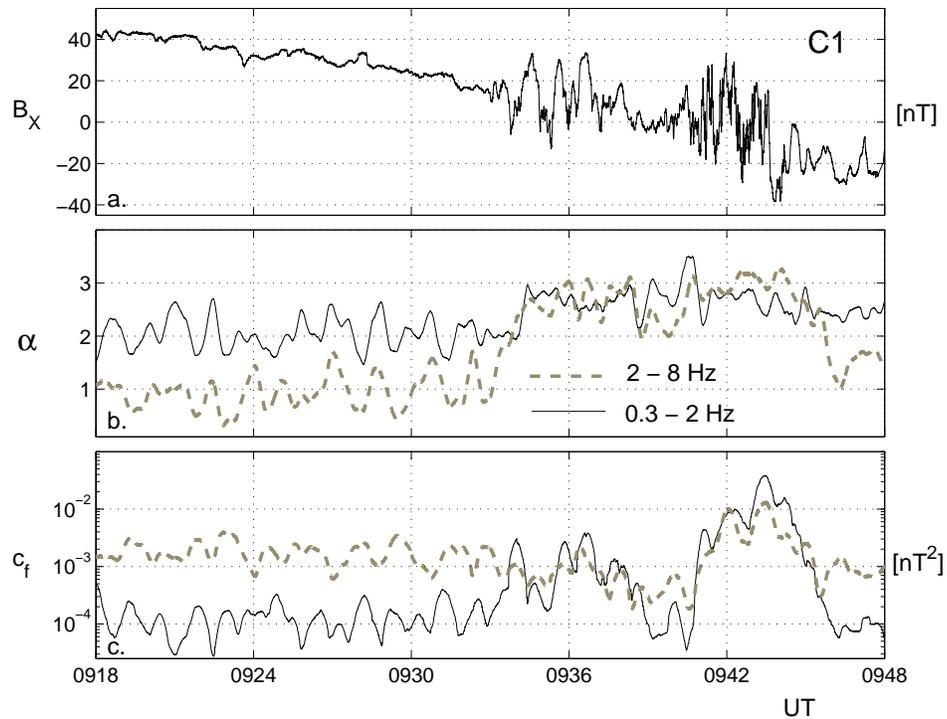}
 \caption{Demonstration of two-scale sliding window analysis.
 (a) $B_X$ magnetic component
 from C1; (b) spectral scaling exponent for $B_X$ - C1;
 (c) spectral power for $B_X$ - C1.
The spectral parameters are estimated within sliding windows of 30s
over frequency ranges 2-8 Hz (dashed grey line) and
 0.3-2 Hz (continuous black line).}
 \end{figure}

 \begin{figure}
\noindent\includegraphics[width=40pc]{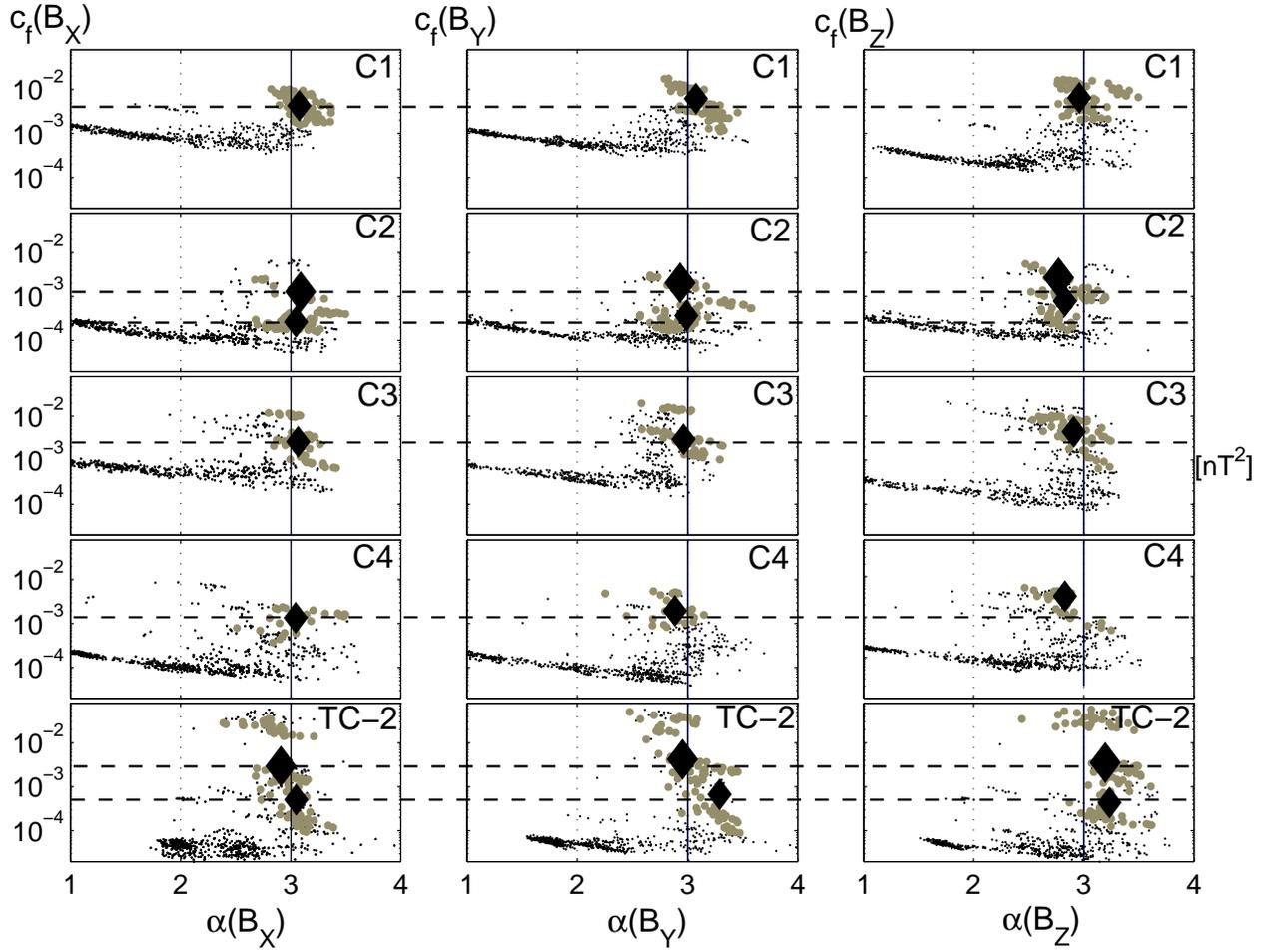}
 \caption{Scatter plots of spectral parameters $c_f(B_i)$ vs
 $\alpha(B_i)$, ($i=X,Y,Z$) for Cluster and TC-2. Small black points
 correspond to noise (see the text); larger grey points represent estimations
 during the intervals depicted by the horizontal dashed lines in Figure 1;
 black diamonds are medians for grey point distributions; better orientation
 is facilitated by vertical lines at $\alpha(B_i)=3$ and dashed horizontal
 lines centered at the medians for $B_X$ components for each spacecraft.}
 \end{figure}

 \begin{figure}
\noindent\includegraphics[width=30pc]{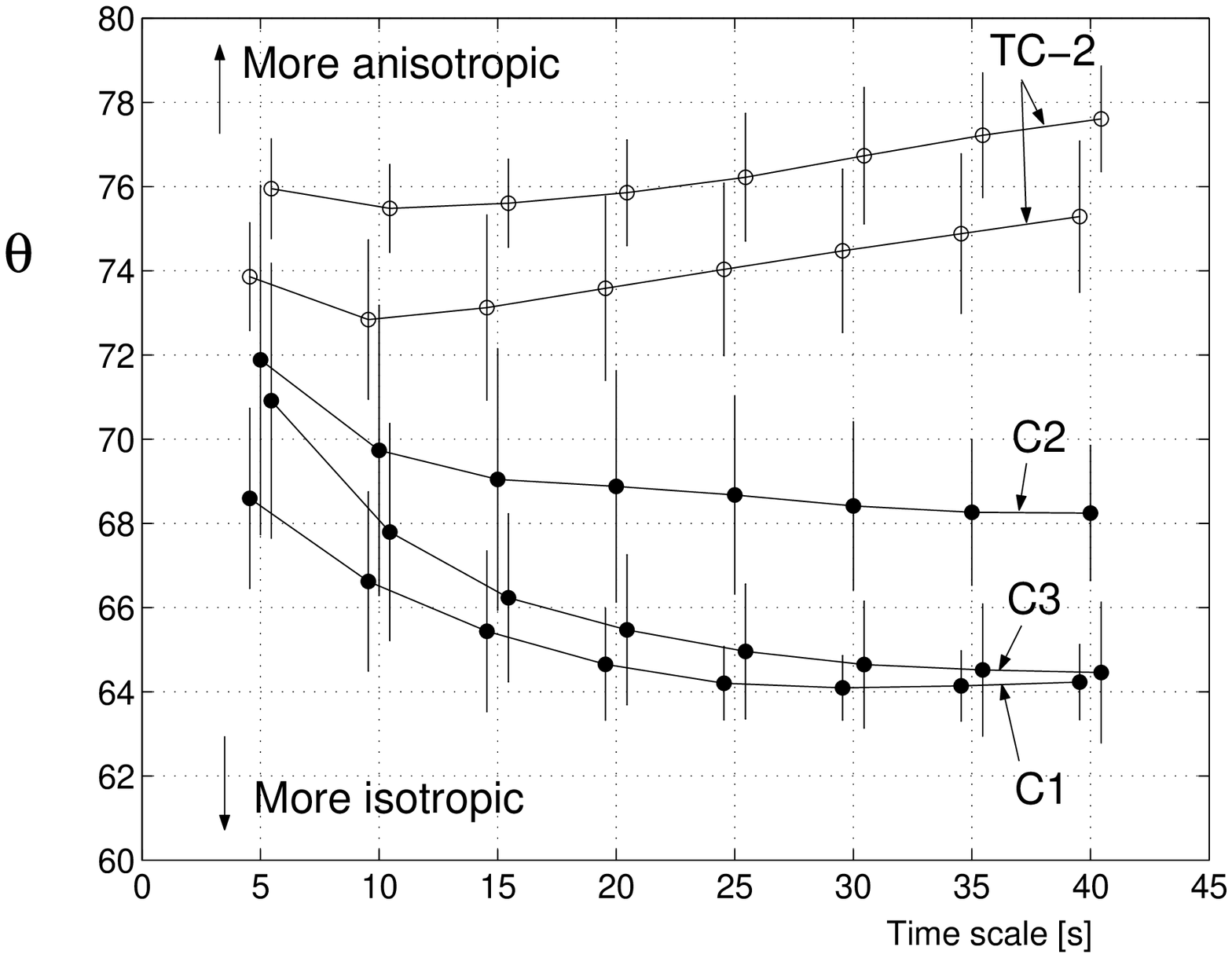}
 \caption{The scale-dependency of anisotropy angle $\theta$
 for validated Cluster and TC-2 data intervals.}
 \end{figure}
\section{Introduction}
Magnetic reconnection (MR) is a fundamental physical process which
was proposed to explain changes in magnetic field topology and
associated conversion of magnetic to kinetic energy in laboratory,
space and astrophysical plasmas [e.g. {\it Birn and Priest}, 2007].
The intensive research in these fields gradually led to a conception
of MR being a multi-scale, non-stationary, 3D process in real
systems. In the terrestrial magnetotail MR occurs in localized
regions, where kinetic effects breaking the frozen-in condition lead
to particle acceleration, plasma bulk flows and other large-scale
outcomes which are back-reacting on proton/electron-scale processes.
There are many unsolved details, even opposing results and
viewpoints about MR and its multi-scale signatures [e.g., {\it
Nakamura et al.}, 2006; {\it Baumjohann et al.}, 2007], therefore
in-situ multi-point measurements are of immense importance. The
four-spacecraft Cluster mission allowed, for example,  for the first
time direct calculations of spatial gradients and unambiguously
observed Hall magnetic fields and current sheet structure near a
reconnection region [{\it Runov et al.}, 2003] and permitted to
estimate the spatial size of bulk plasma flows [{\it Nakamura et
al.}, 2004]. The spatial structure of plasma outflow associated
turbulence was also established using multi-point Cluster
measurements [{\it V\"or\"os et al.}, 2007a]. The multi-point Double
Star (TC) - Cluster spacecraft measurements have already provided
essential information about the large-scale dynamics of magnetotail.
For example, more information was obtained about the propagation of
dipolarization fronts [{\it Nakamura et al.}, 2005; {\it Takada et
al.}, 2006], flow related structures [{\it Volwerk et al.}, 2005,
2007] or the radial extension of flapping motions [{\it Zhang et
al.}, 2005].

In this paper we will analyze and compare the spectral properties of
multi-scale magnetic fluctuations/turbulence associated with the
identified MR events in the companion paper by {\it Sergeev et al.}
(this issue). A strong coupling between reappearing reconnections
and associated turbulence during the September 26, 2005 MR events
has already been emphasized at a qualitative level [{\it Sergeev et
al.}, 2007]. Our goal here is to provide a quantitative description
in terms of scaling properties of non-stationary magnetic
fluctuations within sliding overlapping windows. We also focus on
the anisotropy of fluctuations which is introduced by a local mean
magnetic field. The estimated anisotropies indicate a clear
difference between the physical processes at the Earthward and
tailward outflow sides of MR events at the positions of the TC-2 and
Cluster spacecraft. Variance anisotropy and  scale-dependent
increase of anisotropy at Cluster's location (14-16 Re) indicates
the presence of reconnection outflow associated magnetohydrodynamic
cascading turbulence. At the geostationary distance, where TC-2 is
located (~6.6 Re), variance anisotropy is absent, the anisotropy is
not scale dependent and the fluctuations do not correspond to a
simple cascading turbulence. In this case magnetic fluctuations are
driven differently.

In the analysis presented in this paper  magnetic field data from
the Cluster Flux Gate Magnetometer (FGM, [{\it Balogh et al.},
2001]) and from the Double Star Magnetic Field Investigation (FGM,
[{\it Carr et al.}, 2005]) will be used, both available with 22 Hz
resolution. Spin averaged (4s) bulk velocity data from the Cluster
Ion Spectrometry experiment (CIS, [{\it R\`eme et al.}, 2001]) will
be used for identification of bulk flow associated magnetic
fluctuations at the position of Cluster spacecraft.

\section{Spectral analysis of magnetic fluctuations}
In a companion paper by {\it Sergeev et al.} (this issue), some
signatures and consequences of three intense repeatedly occurring MR
events between 08 and 10 UT on September 26, 2005 at near-Earth
magnetotail are studied using data from Cluster, Double Star (TC-2),
GOES10 and LANL spacecraft.  The MR events were identified on the
basis of simultaneous observations in the inflow and outflow regions
and near-separatrix regions, including energetic electron bursts,
Alfv\`enic outflows, current densities, quadrupole magnetic fields,
electric fields, injections of energetic particles to geostationary
orbit and perturbations of the plasma sheet. We are not going into
all the details here, however, we will accept the working hypothesis
that the observed plasma flows and magnetic fluctuations are
associated with MR events ({\it Sergeev et al.}, this issue).
Magnetospheric modeling and mapping together with the corresponding
auroral brightenings observed by IMAGE auroral imager helped to
identify magnetic field configurations. During this unique period,
the {\it Cluster (14-16 Re) - X-line - TC-2 (6.6 Re)} constellation
and the auroral brightening region were aligned along the same
meridian (~23 h MLT) and Cluster and TC-2 probed opposite sides of
MR associated dynamical activations. Figure 1 shows the GSM
positions of the spacecraft, supposed to be near the central
meridian of repeatedly observed activations. For better orientation,
the approximate position of the X-line is depicted in Figure 1a. The
horizontal dashed lines in Figure 1 c show the GSM Z extent of
reconnection outflow at Cluster positions during the event around
0942 UT [{\it V\"or\"os et al.}, 2007a], (see later). PEACE and CIS
spectrograms (not shown), together with magnetic and plasma
parameters confirm that the Cluster spacecraft stay inside the
plasma sheet during the whole period of study. For example, the
plasma density was about 1 $cm^{-3}$ and the plasma $\beta >$ 1. Ion
moments are not available from TC-2, but the electron moments from
the PEACE instrument and magnetic field variations show that TC-2
was in the middle of the plasma sheet (see {\it Sergeev et al.},
this issue). For example, the electron density was about 1.2 - 1.4
$cm^{-3}$ and the electron temperature varied between 2-3 keV at
TC-2 during the activations.

The two upper subplots in Figure 2a, b show the $B_X$ magnetic
components from the Cluster spacecraft. Figure 2c shows the
available bulk velocity data from the C1 (Hot Ion Analyzer) and C4
(Composition Distribution Function) spacecraft. In Figure 2d the
$B_X$ magnetic component from TC-2 is depicted. The time period from
0930 to 0948 UT in Figure 2 contains the events on September 26,
2005, labeled as b.) and c.) in the paper by {\it Sergeev et al.}
(this issue). The separation  between C3 and C4 in Zgsm direction is
$\sim$ 1000 km, while much smaller in other directions (Figure 1).
Therefore, the increased difference between Bx components measured
by C3 and C4 after 0937 UT is a signature of a thin current sheet,
which is expected to occur near the reconnection region. We will
argue below that the magnetic fluctuations at Cluster and at TC-2
positions are different. In Figures 2a,b,d horizontal dashed lines
show the occurrence of multi-scale (wide-band) fluctuations in
dynamic spectra (not shown). Although the positions of Cluster
spacecraft relative to the neutral sheet or bulk flows are changing
in time, the magnetic fluctuations are very intense during the major
part of the depicted interval. The magnetic fluctuations are
especially enhanced at C1 and C3 (black curves in Figures 2a, b) due
to the spacecraft position being most of the time in the central
plasma sheet. This is in contrast with magnetic fluctuations at TC-2
where the short active periods near 0933 and 0942 UT are separated
by a rather quiet period.

Between $\sim$0932 and 0941 UT the C1 spacecraft observed continuous
fast tailward bulk flow suggesting that the X-line was Earthward of
X = -16 $R_e$. Most of the time C4 was in the boundary plasma sheet
and sporadically detected Earthward flows suggesting that the
X-line, at least sporadically, could appear close to X = -16 $R_e$
(Figure 1a). Later on, between $\sim$0941 and 0943 UT, both C1 and
C3 observe strong tailward flow reaching -1000 km/s. This can have
different explanations, e.g. Earthward moving X-line, MR active in a
different location or even a complex 3D geometry. The two vertical
dashed lines in Figure 2 show the beginning of firstly observed
multi-scale magnetic fluctuations by Cluster at 093145 UT (C2) and
at 094030 UT (C1, C3). The associated beginnings of multi-scale
magnetic fluctuations at TC-2 position are time-delayed by
approximately 15-20 s. Though, full-scale substorms were not
observed, the above time delays together with the observed auroral
brightenings and injections of energetic particles to geostationary
orbit indicate that, the activation events can be described in terms
of the Near Earth Neutral Line (NENL) model (see {\it Sergeev et
al.}, this issue).

In summary, the length (horizontal dashed lines in Figure 2) and the
beginnings and ends of multi-scale magnetic fluctuations at Cluster
and TC-2 positions are different and the magnetic X-line(s) seems to
be closer to the Cluster position (Figure 1a). Therefore, it is not
a priori clear how to compare straightforwardly the spectral
properties of fluctuations at the positions of Cluster and TC-2, or
how long and which data intervals (physical processes) should be
considered in intercomparison. Moreover, the physical processes in
the plasma sheet are multi-scale and non-stationary [{\it V\"or\"os
et al.}, 2006]. A possible solution for this problem is the
estimation of spectral properties in sliding overlapping windows.
The length of the sliding window $W$ has to be chosen according to
the available length of - at least quasi-stationary - physical
processes. Contrarily, to be able to estimate the spectral
properties over multiple scales, $W$ should be as large as possible.
Since magnetic data from Cluster and TC-2 are available with the
same resolution (22 Hz), $W$ will be equal for all spacecraft.
Obviously, $W$ cannot be larger than the lifetime of fluctuations we
are interested in, e.g. the length of the first activation measured
by TC-2 before 0933 UT is about 1 min. The second activation at TC-2
after 0941 UT shows visually a longer duration ($\sim$ 3 min), but
one cannot be sure if the process is really stationary during the
whole activation interval. The multi-scale magnetic fluctuations
e.g. at C1 or C3 after 0941 UT persist also a few minutes. Similar
fluctuations in the central plasma sheet, however, are known to be
driven by plasma bulk flows and those are bursty and time dependent
[{\it Baumjohann et al.}, 1990; {\it Angelopoulos et al.}, 1992;
{\it V\"or\"os et al.}, 2004a, b]. For example, the bulk flows
exhibit $\sim$ -1000 km/s speed no longer than a minute around 0942
UT, then the speed is decreasing (Figure 2c). Therefore, we choose
$W =$ 30 s and shift the window by 2 s along the time series after
each calculation step.

Within each analyzing window the spectral scaling parameters $c_f$
and $\alpha$ in the relation $P(f) \sim c_f f^{-\alpha}$ are
estimated, using a wavelet technique proposed by {\it Abry et al.},
[2000], where $f$ is the frequency, $\alpha$ represents the scaling
index and $c_f$ is a nonzero magnitude parameter. The method was
adapted for the analysis of plasma sheet turbulence by {\it
V\"or\"os et al.} [2004b]. Here, we recall only the main steps of
the wavelet technique. First, a discrete wavelet transform of the
data is performed over a dyadic grid $(scale, time)=(2^j,2^j t)$ and
$j,t \in \bf{N}$. Then, at each octave $j=log_2 2^j$, the variance
$\mu_j$ of the discrete wavelet coefficients $d_x(j,t)$  is computed
through $ \mu_j = \frac{1}{n_j}\sum_{t=1}^{n_j}d_x^2(j,t) \sim
2^{j\alpha}\ c_f$, where $n_j$ is the number of coefficients at
octave $j$. Finally, $\alpha$ is estimated by constructing a plot of
$y_j\equiv log_2 \mu_j$ versus $j$ (logscale diagram) and by using a
weighted linear regression over the region $(j_{min}, j_{max})$
where $y_j$ is a straight line. $c_f$ represents the intercept of
that line with $j=0$ and has the dimension of variance ($nT^2$ in
our case).

Having the window $W$ and the resolution fixed, the scaling
properties of fluctuations can only be studied over the time scales
of $\sim$ 0.1 - 5 s (0.2 - 10 Hz) (with 95\% confidence). We
consider that, magnetic fluctuations are statistically stationary
only if the estimated spectral characteristics are similar at least
over 2 window length (2W = 1 minute), consisting 30 analyzing
windows having the step of 2 s. Physical processes with shorter
duration than 2W cannot be resolved by this method and the sliding
window analysis produces fluctuating estimates of $c_f$ and
$\alpha$, usually underestimating $\alpha$. The estimations are also
influenced by the $\sim$ 4 s spin tone. This effect was partially
reduced by careful calibration and by examination of time scales
below 4 s.

First, we demonstrate, how multi-scale fluctuations behave over two
nearby scales during active and quiet periods. For comparison with
the more standard Fourier approach we transformed the wavelet based
scales into the approximate Fourier frequencies. The relationship
between wavelet scales and Fourier frequencies can be derived by
finding the wavelet transform of a pure cosine wave with a known
Fourier frequency (e.g. [{\it Torrence and Compo}, 1998]). Figure 3
shows the comparison of power spectra estimated from TC-2 data
during a quiet interval (interval (a): the analyzing window W is
centered at 0931 UT, see Figure 2d) and during reconnection
associated fluctuations (interval (b): W is centered at the first
activation before 0933 UT). During the quiet period (interval (a))
two neighbouring frequency ranges 0.3 - 2 Hz (larger scale) and 2 -
8 Hz (smaller scale) show scalings with different scaling exponents.
However, no spectral break is observed over the same frequency
ranges during reconnection associated magnetic fluctuations
(interval (b)) and the power of the fluctuations is larger than
during the quiet period. It indicates that developed multi-scale
fluctuations are present, driven by the reconnection outflow.

Figure 4 demonstrates the sliding window estimation of scaling
parameters $\alpha$ (Figure 4b) and $c_f$ (Figure 4c) for magnetic
field $B_X$ components from C1 (Figure 4a). A longer interval
between 0918 and 0948 UT is shown to see the difference between
quiet and active periods. Again, both scaling parameters are
estimated over the two neighbouring frequency ranges 0.3 - 2 Hz
(larger scale, continuous black curves) and 2 - 8 Hz (smaller scale,
dashed grey curves). When magnetic fluctuations are rare or absent
(0918 - 0933 UT) the scaling parameters are fluctuating near the
noise level and the difference of estimations between neighbouring
scales is large. Between 0933 and 0938 UT only the large-scale power
is enhanced over the noise level, when $B_X$ exhibits low-frequency
flapping-like motion almost without higher-frequency activity.
Between 0941 and 0945 the burst plasma flow related multi-scale
magnetic fluctuations lead to enhanced power and closer scaling
index estimations over both frequency ranges. It indicates again
that multi-scale flow-driven turbulent fluctuations are present (see
further details in {\it V\"or\"os et al.} [2004b, 2006]).

Figure 5 shows the scatter plots of $\alpha (B_i)$ vs. $c_f(B_i)$,
computed over frequency range 0.3 - 8 Hz, for all magnetic
components (i= X, Y, Z). The top three rows contain Cluster and the
bottom row TC-2 spectral estimations. Three populations of data
points are well discernible. The majority of small black points
showing large spread in $\alpha$ with the smallest powers $c_f$
organized into a narrow arm-like structures correspond to noise. As
it was already mentioned in two-scale analysis, no activity or
scale-restricted fluctuations results in noisy or/and small power
estimations. The larger grey points with higher powers concentrating
between $\alpha \in$ (2.5-3.5) correspond to multi-scale
fluctuations/turbulence. The remaining black point populations are
due to noisy transients, short duration events, etc. The black
diamonds show the medians of grey point distributions (multi-scale
turbulence). C2 and TC-2 spacecraft data contain 2 periods with
multi-scale turbulence resulting in two distributions of grey points
and two median values. The larger diamonds denote later time
intervals. We will not analyze the differences between the different
activation periods because the corresponding grey point
distributions are overlapping (no additional color needed). Instead,
we will concentrate on more pronounced differences between Cluster
and TC-2 magnetic fluctuations.

To allow better comparisons between the spectral parameters for each
magnetic component and spacecraft, 3 vertical lines are drawn at
$\alpha(B_i)=3$. The dashed horizontal lines are centered at median
values (black diamonds) for $B_X$ components (in first column of
Figure 5). For $B_X$ and $B_Y$ components the median values of
scaling indices are rather similar, near $\alpha \sim 3$ for all
spacecraft. The median values of $\alpha$ are systematically below 3
for Cluster $B_Z$ and over 3 for TC-2 $B_Z$ components. This
component dependent difference in $\alpha$ can be explained by
stronger $B_Z$, namely more dipolar field at the position of TC-2
and by peculiarities of the interaction of plasma flows with the
dipolar field. For Cluster data (top four rows in Figure 5), the
median values of powers $c_f(B_i)$ associated with multi-scale
fluctuations (black diamonds) show a weak but consistent increase
when going from $B_X$ to $B_Y$ and $B_Z$. At the same time the
corresponding scaling indices $\alpha (B_i)$ change from $\sim$ 3.1
to $\sim$ 2.9. This corresponds to variance anisotropy in plasma
sheet turbulence where perpendicular variances relative to the mean
magnetic field are larger than the parallel one [{\it V\"{o}r\"{o}s
et al.}, 2004b]. In Cluster position the variances are increased in
the GSM Y, Z components, because the mean field direction is roughly
in GSM X direction in a stretched plasma sheet. Variance anisotropy
occurs in compressible MHD, but the controlling factor is mainly the
plasma $\beta$ and not the proximity to the incompressible state
[{\it Matthaeus et al.}, 1996]. It is also known from 3D MHD
simulations of turbulence that, for $\beta \sim 1$ variance
anisotropy is significant, however, over $\beta > 40 $ the
anisotropy disappears [{\it Matthaeus et al.}, 1996]. The power of
compressional fluctuations over MHD scales is known to be larger
than the power of right-hand and left-hand polarized waves in the
near-Earth plasma sheet [{\it Volwerk et al.}, 2004]. More
importantly, plasma $\beta$ varied between 1 and 10 during the
activation periods, which can explain the appearance of variance
anisotropy at the position of Cluster spacecraft.

The spread around the median powers (larger grey points) is much
larger at the position of TC-2 than of Cluster. The maximum values
of powers are close to $10^{-1}$ $nT^2$ at TC-2 for all magnetic
components. As described above, TC-2 is in the plasma sheet during
the activations, presumably crossing regions with plasma $\beta>1$,
but still in a region with favorable conditions for the occurrence
of variance anisotropy (plasma $\beta$ certainly will not approach
40). Nevertheless, variance anisotropy is not observed at the
position of TC-2 (bottom row in Figure 5). This may indicate that
the magnetic fluctuations over the considered scales might be driven
by different physical processes at Cluster and TC-2 positions. In
order to investigate further this possibility we will analyze the
anisotropy of fluctuations in wave number space. In the presence of
sufficiently strong mean magnetic field ($B_0$) scale-dependent
spectral anisotropy in wave vector space is dynamically and robustly
generated in MHD cascading turbulence [{\it Shebalin et al.}
[1983]]. This offers the possibility to compare the expected mean
field influenced spectral properties of MHD turbulence with the
observed multi-spacecraft data. We note, that the variance
anisotropy is not necessarily the same as the anisotropy in wave
vector space, because there is no full correspondence between the
most energetic wave vectors and the minimum variance directions
[{\it Matthaeus et al.}, 1996].

\section{Scale-dependent anisotropy}
{\it Shebalin et al.} [1983] studied incompressible MHD anisotropies
arising in wave vector space in the presence of a mean magnetic
field. They studied the interaction of opposite-traveling wave
packets and found that in wave vector space, those interactions
produce modes with wave vectors preferentially perpendicular to the
mean magnetic field. {\it Goldreich and Sridhar} [1995] proposed a
balance condition between parallel and perpendicular modes
(parallel-propagating Alfven waves and perpendicular eddy motions).
On this basis it was shown that a scale-dependent anisotropy
appears, namely $k_{||} \sim k_{\perp}^{2/3}$ (or $l_{||} \sim
l_{\perp}^{2/3}$, where $l \sim 1/k$ is the characteristic scale of
eddies in parallel and perpendicular directions), which means that
the anisotropy is increasing with decreasing scale.

As a quantitative measure of spectral anisotropy, it is convenient
to use the anisotropy angles, $\theta$ introduced by {\it Shebalin
et al.} [1983]
\begin{equation}
tan^2 \theta = \frac{<k^2_{\perp}>}  {<k^2_{||}>}
\end{equation}
To isotropic fluctuations corresponds $\theta \sim 54^{\circ}$,
purely parallel (slab) fluctuations exhibit $\theta =0^{\circ}$, and
fully perpendicular (2D) fluctuations results in
$\theta=90^{\circ}$. Now, the question is how to make measurements
in the wave vector space?

Based on four-point Cluster measurements the so-called k-filtering
technique was proposed to estimate the magnetic field energy
distribution in both the angular frequency and wave vector spaces
[{\it Sahraoui et al.}, 2003]. In our case, this technique cannot be
used, because there are no four identical spacecraft at the position
of TC-2. Also, the Cluster spacecraft are frequently merged into
different physical regions in the dynamical magnetotail [{\it
V\"or\"os et al.}, 2007a], the k-filtering technique, however,
assumes homogeneous and stationary wave-fields in space.

{\it Oughton et al.} [1998] presented a simple model for the scaling
of the anisotropy as a function of the fluctuating magnetic field
$\delta b$ over the local mean magnetic field $B_0$. On the basis of
theoretical arguments and numerical simulations they have shown that
the anisotropy angle can be estimated from the relationship
\begin{equation}
cos^2 \theta = m {\Big\lgroup\frac{\delta b}  {B_0 + \delta
b}\Big\rgroup}^2 +c
\end{equation}
we use the values $m = 0.4$ and $c=0.01$ obtained from Figure 3 in
[{\it Oughton et al.}, 1998]. Both $m$ and $c$ can change with Mach
number and Reynolds number, but small changes of these numbers do
not influence significantly our results. We note that by using the
above relationship the anisotropy angle defined in the Fourier space
can be estimated in the time-domain. It was also shown that the
approximate relationship is valid only for some intermediate values
\begin{equation}
0.1 \le \frac{\delta b} { (B_0 + \delta b)} < 1
\end{equation}
For the estimation of $\theta$ we used the same sliding window
technique as above. The mean magnetic field is $B_0^2(t,W)=\langle
B_X^2(t) + B_Y^2(t) + B_Z^2(t)\rangle$ and $\langle \rangle$ stands
for averaging over the window $W$. The fluctuations $\delta b$ were
obtained within each $W$ through $\delta b^2(t,W) = \langle
(B_X-\langle B_X\rangle )^2\rangle + \langle (B_Y-\langle B_Y\rangle
)^2 \rangle + \langle (B_Z-\langle B_Z\rangle )^2\rangle$. The
time-scale in this time-domain analysis is represented by $W$. We
note, that the mean magnetic field within the windows W was always
different from zero, even if locally $B_X \sim 0$ nT can occur (e.g.
at Cluster around 0942 UT, Figure 2). One can perform sliding window
analysis estimating $\theta$ for different $W$ (different scales)
keeping the time shift unchanged. Interestingly, when multi-scale
fluctuations were absent the ratio $\delta b / (B_0 + \delta b)<0.1$
indicating that $\theta$ has no physical meaning during quiet
periods (not shown). For proper steady estimations of $\theta$
similar values of $\delta b / (B_0 + \delta b)$ were required at
least over 2 window length.

Out of seven data intervals of multi-scale fluctuations (horizontal
dashed lines in Figure 2 and diamonds in Figure 5), three data
intervals from Cluster and two data intervals from TC-2 were found
for which $\theta$ could be estimated according to Equations 2 and
3. During the remaining two intervals at TC-2 and C4 near 0942 UT,
the estimations of $\theta$ were not stationary enough. Figure 6
shows the scale dependence of estimated anisotropy angles $\theta$.
The error bars represent standard deviations. Turbulence at the
position of Cluster exhibits scale-dependent anisotropy. The
anisotropy angle is increasing towards small scales, but below the
time scale of 10-15 s the anisotropy angle is almost the same for
TC-2 and Cluster spacecraft. At the TC-2 position the fluctuations
are more anisotropic, but almost scale independent, exhibiting a
weak increase towards larger time scales.

Both variance anisotropy and scale-dependent anisotropy in wave
vector space seem to confirm the above supposition that the
multi-scale magnetic fluctuations at the positions of Cluster and
TC-2 spacecraft are different.

\section{Conclusions}
In this paper magnetic fluctuations at Cluster and Double Star were
compared using spectral and time-domain methods. The comparison is
facilitated because of the magnetometers with similar technical
parameters on these spacecrafts [{\it Balogh et al.} 2001; {\it Carr
et al.} 2005].

The spectral power increased during the periods of multi-scale
fluctuations. The power of fluctuations over frequency range 0.3 - 8
Hz are 1-2 orders of magnitude above the quiet time periods at
Cluster positions (14-16 $R_e$, Figures 4, 5). This is an
established feature of bursty bulk flow associated magnetic
fluctuations [e.g. {\it V\"or\"os et al.}, 2004b]. In fact, the
multi-scale magnetic fluctuations at Cluster are all flow related
(the five intervals labeled by horizontal dashed lines in Figures
1a, b). The ion data is not available from C2, but the GSM Z
distance between C1 and C2 was only about 2000 km, C1 observed a
tailward bulk flow and C2 was closer to the neutral sheet than C1.
The plasma flow origin of magnetic fluctuations observed by Cluster
after 0941 UT was demonstrated by {\it V\"or\"os et al.} [2007a].
The power of fluctuations are 1-4 orders of magnitude above the
quiet time periods at TC-2 position ($\sim 6.6 R_e$, Figure 5,
bottom panels). The TC-2 results can be compared with the Charge
Composition Explorer (CCE) of the Active Magnetospheric Particle
Tracer Explorer (AMPTE) observations of magnetic fluctuations in the
pre-midnight and post-midnight sectors at radial locations of 7.4 -
8.8 $R_e$. During current disruption events observed by AMPTE wave
powers of 2-3 orders of magnitude above the pre-onset period were
measured over the frequency ranges of 0.1 - 4 Hz [{\it Lui et al.},
1992]. The scaling index at AMPTE is changing between $\alpha = 2.0
- 2.6$ [{\it Lui et al.}, 1992; {\it Ohtani et al.}, 1995]. The
spectral scaling index at TC-2 and Cluster spacecraft changed
between $\alpha = 2.5 - 3.5$ , showing a small, presumably $B_Z$
component related difference, but negligible between the events near
0933 and 0942 UT (small and large diamonds in Figure 3). Having in
mind different magnetometers, noise levels and slightly different
frequency ranges in TC-2 and AMPTE spacecraft, the estimated
spectral parameters are rather close to each other. The differences
in spectral parameters can also appear because of the different
positions of spacecraft, observing fluctuations in regions with
different plasma $\beta$, proton gyroperiod [{\it Ohtani et al.},
1995] or boundary effects [e.g. {\it Borovsky and Funsten}, 2003].
Notice, that the observed spectral indices are different from the
values of 3/2 or 5/3 which correspond to hydrodynamic or
magnetohydrodynamic inertial range scalings, respectively. It was
shown recently by {\it V\"or\"os et al.} [2007b] that, the inertial
range can be identified only when at least several minute long
quasi-stationary plasma flows exist in the turbulent plasma sheet.
For frequencies larger than the proton gyrofrequency (typically
0.1-0.4 Hz in our case), the spectrum steepens, exhibiting spectral
indices near 3. This steepening cannot be caused by dissipation via
kinetic wave damping, because it would result in a strong cutoff in
the power spectra rather than a power law (Li et al., 2001). When
the energy is not dissipated it has to be transferred further
towards higher frequencies. The spectral energy transfer in the
small-scale cascade might be controlled by other physical processes
exhibiting different characteristic time scales than the inertial
range magnetohydrodynamic turbulence. The physical mechanisms
involved in the generation of this high frequency scaling range can
be e.g. Hall physics related in the presence of dispersive effects
[e.g. {\it Alexandrova et al.}, 2007]. The small-scale cascade is
intermittent [e.g. {\it Alexandrova et al.}, 2007; {\it V\"or\"os et
al.}, 2003, 2007], which indicates that intermittency is not
restricted to the inertial range [{\it Bruno and Carbone}, 2005].
Numerical simulations of MHD turbulence show that neither the
magnetic field anisotropy is restricted to the inertial range and
the small scale cascade continues to be anisotropic [e.g. {\it Cho
et al.}, 2002].

These findings, in agreement with earlier results, indicate that
fluctuations/turbulence exhibit different spectral characteristics
in the tailward (Cluster)/Earthward (TC-2) reconnection outflow
regions, substantiated further by the fact that the consequences of
the same reconnection events are observed by very similar
magnetometers. The fluctuations at Cluster and TC-2 positions might
be driven by different physical processes, because the physical
situations are different. Physically, in the tailward outflow region
(Cluster) the strong plasma flow drives the magnetic field
fluctuations, the latter representing a passive scalar field [{\it
V\"or\"os et al.}, 2004a]. In the Earthward outflow region (TC-2)
$B_Z$ is stronger, the magnetic field is more dipolar, the leading
processes are flow breaking [e.g., {\it Shiokawa et al.}, 1997],
dipolarization [e.g., {\it Baumjohann et al.}, 1999] or tail current
disruption [e.g., {\it Lui et al.}, 1992]. Unlike the reconnection
outflows at the magnetopause, Earthward - tailward plasma outflows
are not symmetric in the tail. On the Earthward side the flows are
more affected by the obstacle represented by the magnetic wall - a
stronger near-Earth magnetic field. The reconnection outflow
asymmetry in the plasma sheet has already been observed, e.g. in
connection with slow-mode shocks and the Wal\'en test [{\it Eriksson
et al.}, 2004].

In summary, the difference between plasma flow driven and stronger
dipolar field dominated fluctuations is discernible in both variance
and scale dependent spectral anisotropy features. There are
favorable conditions in both locations for variance anisotropy, but
it appears only at Cluster positions. Scale dependent anisotropy in
wave-vector space is present also only at Cluster locations, though
TC-2 observes a slightly larger anisotropy angles. Theory and
simulations of magnetohydrodynamic cascading turbulence with
magnetic field and not too high plasma $\beta $ predict both types
of anisotropies. We can speculate that the lack of anisotropy in
TC-2 location can be explained through: (a) magnetic fluctuations
which are not associated by a turbulent cascade; for example
currents can heat the plasma over small spatial scales, in which
case substantial part of energy would not be transferred through a
turbulent cascade; (b) magnetic fluctuations which are associated by
a turbulent cascade, but without scale dependent anisotropy. The
anisotropy level can be higher, but saturated because of turbulence
ageing or/and mixing and flow breaking near the more dipolar field.
As a matter of fact, the largest anisotropy angle at the Cluster
position (which is closer to the reconnection site) is roughly the
same as the average anisotropy angle at TC-2 position. In any case,
flow breaking/mixing due to the near-Earth magnetic obstacle and the
associated plasma/current instabilities can explain the lack of
anisotropies at TC-2 position.

%% ------------------------------------------------------------------------ %%
%
%  ACKNOWLEDGMENTS
%
%% ------------------------------------------------------------------------ %%
\begin{acknowledgments}
The authors thank H. U. Eichelberger for great help in the data
processing. The work by Z. V\"or\"os was partially supported by the
Grant Agency of the Academy of Sciences of the Czech Republic (Grant
No.: B300420509). Part of the work by M. Volwerk was financially
supported by the German Bundesministerium f\"ur Bildung und
Forschung and the Zentrum f\"ur Luft- und Raumfahrt under contract
50 OC 0104.
\end{acknowledgments}

%% ------------------------------------------------------------------------ %%
%
%  REFERENCE LIST AND TEXT CITATIONS
%
%% ------------------------------------------------------------------------ %%
%
% If you use BiBTeX for your References, please do not send
% your bibliography database. Copy the reference list
% from your .bbl file into your article file before submission:
%
%1. Run LaTeX on your LaTeX file.
%
%2. Run BiBTeX on your LaTeX file.
%
%3. Open the new .bbl file containing the reference list and
%copy all the contents into your LaTeX file after the
%acknowledgments section;
%
%4. Comment out the old \bibliographystyle and \bibliography commands.
%
%5. Run LaTeX on your new file before submitting.

%Failure to follow these instructions will require manual
%intervention through hard keying of information,
%which can introduce errors.

\end{article}

\end{document}